\documentclass[aps,pra,superscriptaddress,twocolumn,citeautoscript,10pt, rsi,reprint]{revtex4-1}
\usepackage{amsmath}
\usepackage{graphicx}
\usepackage{bm}
\usepackage{soul}
\usepackage{xcolor}
\usepackage{amsmath}
\usepackage{float}
\usepackage{comment}
\usepackage{graphicx}
\usepackage{dcolumn}
\usepackage[colorlinks=true,bookmarks=false,citecolor=blue,linkcolor=blue,hyperfootnotes=true,urlcolor=blue]{hyperref}

\begin{document}
\title{ First-principles study of electron transport in ScN}
\author{Sai Mu}
\affiliation{Materials Department, University of California, Santa Barbara, California, 93106-5050, USA}
\author{Andrew J. E. Rowberg}
\affiliation{Materials Department, University of California, Santa Barbara, California, 93106-5050, USA}
\author{Joshua Leveillee}
\affiliation{Oden Institute for Computational Engineering and Sciences, The University of Texas at Austin, Austin, Texas 78712, USA}
\affiliation{Department of Physics, The University of Texas at Austin, Austin, Texas 78712, USA}
\author{Feliciano Giustino}
\affiliation{Oden Institute for Computational Engineering and Sciences, The University of Texas at Austin, Austin, Texas 78712, USA}
\affiliation{Department of Physics, The University of Texas at Austin, Austin, Texas 78712, USA}
\author{Chris G. Van de Walle}
\affiliation{Materials Department, University of California, Santa Barbara, California, 93106-5050, USA}

\begin{abstract}
We investigate the conduction-band structure and electron mobility in rocksalt ScN based on density functional theory.
The first-principles band structure allows us to obtain band velocities and effective masses as a function of energy.
Electron-phonon scattering is assessed by explicitly computing the $q$-dependent electron-phonon matrix elements, with the inclusion of the long-range electrostatic interaction.
The influence of free-carrier screening on the electron transport is assessed using the random phase approximation.
We find a notable enhancement of electron mobility when the carrier concentration exceeds 10$^{20}$ cm$^{-3}$.
We calculate the room-temperature electron mobility in ScN to be 587 cm$^2$/Vs at low carrier concentrations.
When the carrier concentration is increased, the electron mobility starts to decrease significantly around $n=10^{19}$ cm$^{-3}$, and drops to 240 cm$^2$/Vs at $n=10^{21}$ cm$^{-3}$.
We also explore the influence of strain in (111)- and (100)-oriented ScN films.
For (111) films, we find that a 1.0\% compressive epitaxial strain increases the in-plane mobility by 72 cm$^2$/Vs and the out-of-plane mobility by 50 cm$^2$/Vs.
For (100) films, a 1.0\% compressive epitaxial strain increases the out-of-plane mobility by as much as 172 cm$^2$/Vs, but has a weak impact on the in-plane mobility.
Our study sheds light on electron transport in ScN at different electron concentrations and shows how strain engineering could increase the electron mobility.

\end{abstract}
\maketitle
\section{\label{sec:intro}Introduction}
ScN is a semiconducting transition-metal nitride that has received significant attention due to the demonstrated synthesis of high-quality thin films \cite{oshima2014hydride,ohgaki2013electrical,saha2013electronic} and their potential applications in thermoelectric \cite{zide2005thermoelectric,saha2012thermoelectric,kerdsongpanya2017phonon} and electronic \cite{biswas2019development} devices.
The material exhibits high mobility and carrier concentrations \cite{al2020properties,deng2015optical} and ambipolar dopability \cite{nayak2019rigid}.
Crystallizing in the rocksalt structure, ScN can be integrated with several technologically important III-nitride semiconductors, 
leading to reduced dislocation density with the use of a GaN epilayer~\cite{johnston2009defect}, and giant interfacial carrier densities in ScN/GaN heterostructures \cite{adamski2019giant}.

The electronic structure of ScN has historically been a topic of debate \cite{biswas2019development}.
Early reports suggested that it might be a semimetal \cite{neckel1975results,monnier1985electron}, with the conduction-band minimum (CBM) at X lower in energy than the valence-band maximum (VBM) at $\Gamma$, but it is now understood that ScN is an indirect-band gap semiconductor.
Experimental values for the fundamental band gap have been reported ranging from 0.9 to 1.3 eV \cite{al2004surface, gall2001electronic,saha2013electronic,deng2015optical}, while
computational studies have produced values between 0.8 and 1.1 eV \cite{saha2017compensation,qteish2006exact,stampfl2001electronic,lambrecht2000electronic}.
A more rigorous study of the electronic structure is still lacking, particularly relating to how band curvature, which governs velocities and effective masses, changes with carrier energy.

A high carrier mobility is desirable for many of ScN's applications.
In thermoelectric energy conversion devices, ScN's high mobility contributes to a higher figure of merit \cite{kerdsongpanya2017phonon,burmistrova2013thermoelectric}.
Similarly, ScN integration with GaN-based electronic or optoelectronic devices is made attractive by the high electron mobility \cite{moram2010low,johnston2009defect}. 
It is thus crucial to understand the factors that impact the mobility, as well as how it varies as a function of carrier concentration.
It is possible to achieve $n$-type doping through the use of O$_\mathrm{N}$ or F$_\mathrm{N}$ donors \cite{kumagai2018point,cetnar2018electronic}.
Reported measurements of room-temperature electron mobility in ScN cover a wide range, from 1 to 284 cm$^2$/Vs \cite{oshima2014hydride,saha2017compensation, saha2013electronic, dismukes1972epitaxial,burmistrova2013thermoelectric, ohgaki2013electrical, biswas2019development}.
The highest reported value, 284 cm$^2$/Vs, was obtained
at a carrier concentration of 3.7$\times$ 10$^{18}$ cm$^{-3}$ in ScN films grown on sapphire by hydride vapor phase epitaxy \cite{oshima2014hydride}.

Accurate values for effective mass, as well as knowledge of how these masses change with carrier concentration, are essential for understanding mobility.
A wide range of electron effective masses has been reported, with experimental measurements for transport effective masses ranging between 0.1 and 0.4 $m_0$~\cite{harbeke1972electron,deng2015optical}, and results from computations ranging from 0.18 to 0.35 $m_0$~\cite{qteish2006exact,deng2015optical}.
These numbers reflect effective masses at the CBM; at high carrier concentrations, the masses can change significantly.

In this paper we report the results of first-principles calculations based on density functional theory (DFT) \cite{hohenberg1964,kohn_self-consistent_1965} and density functional perturbation theory (DFPT) to obtain a comprehensive picture of the electronic structure and electron transport in ScN.
We use a hybrid functional \cite{heyd_hse,HSE06} combined with Wannier interpolation \cite{Pizzi2019wannier90} to obtain an accurate description of the band structure.
This allows us to determine band velocities and effective masses as a function of electron energy, and also to locate inflection points.
Our approach provides accurate values that take the significant degree of anharmonicity of the electronic band structure into account.
We also use the Wannier-interpolated density of states (DOS) and Fermi-Dirac statistics to determine the carrier concentrations for holes and electrons as a function of Fermi level.

We also perform detailed calculations of the electron-phonon (el-ph)-limited electron mobility.
At room temperature, the electron mobility is limited primarily by interactions with the longitudinal optical phonon mode that generates a long-range electric field.
Properly treating this long-range interaction is crucial.
We also include the effects of free-carrier screening on electron mobility, by screening the el-ph matrix elements using the Lindhard dielectric function calculated within the random phase approximation (RPA) for the homogeneous electron gas \cite{Lindhard,hedin1965new}.
We present a detailed analysis of scattering rates and contributions from different phonon modes, and analyze the dependence of mobility on carrier concentrations.

Additionally, we study the impact of strain on mobility.
Thin films of ScN have been grown on MgO \cite{al2000molecular,saha2017compensation,ohgaki2013electrical,burmistrova2013thermoelectric}, sapphire \cite{dismukes1972epitaxial,oshima2014hydride}, and Si \cite{moram2006young} substrates.
Since the lattice mismatch is large, the resulting films are likely relaxed, but residual strain may be present~\cite{saha2017compensation, moram2006young}.
A smaller lattice mismatch occurs for (111)-oriented ScN grown on the $c$ plane of wurzite GaN ($+$0.2\%) and AlN ($-$2.3\%), making pseudomorphic growth plausible.
Since strain is likely to be present in ScN films and is known to affect the electronic structure and carrier mobility of semiconductors \cite{ponce2019route}, we will explore its impact on electron transport.

The paper is organized as follows.
Section~\ref{sec:method} describes the computational methodology and theoretical background.
The main results are summarized in Sec.~\ref{sec:result}, including the electronic (Sec.~\ref{sec:bands}) and phonon band structure (Sec.~\ref{sec:phononbands}) of bulk ScN, electron concentrations as a function of Fermi level (Sec.~\ref{sec:concen}), electron-phonon coupling matrix elements (Sec.~\ref{sec:matrix}), electron scattering rates and their phonon-mode decomposed contributions (Sec.~\ref{sec:rates}), room-temperature electron mobility at various carrier concentrations (Sec.~\ref{sec:mob}), and the effect of screening (Sec.~\ref{sec:screen}).
The strain dependence of electron mobility at low carrier concentrations is investigated in Sec.~\ref{sec:strain}).
Section~\ref{sec:conc} concludes the paper.

\section{Methodology} \label{sec:method}
\subsection{Computational Details}
\label{sec:compdet}

We conduct first-principles calculations based on DFT \cite{hohenberg1964,kohn_self-consistent_1965} and the Heyd, Scuseria, and Ernzerhof (HSE) \cite{heyd_hse,HSE06} screened hybrid functional with 25\% mixing of exact exchange, as implemented in the Vienna \textit{Ab initio} Simulation Package (VASP) \cite{Kresse1993,Kresse1996}.
We use an energy cutoff of 520 eV for the plane-wave basis set, and the core electrons are described with projector-augmented-wave potentials \cite{Blochl_paw1,Kresse_paw2} with the Sc $4s^{2}$ $3d^{1}$ and N $2s^{2}$ $2p^{3}$ electrons treated as valence.
We integrate over the Brillouin zone (BZ) using a 10 $\times$ 10 $\times$ 10 $k$-point grid.
The HSE calculated equilibrium lattice parameter is 4.48 {\AA}, very close to the experimental value of 4.50 {\AA} \cite{travaglini1986electronic}.
To enable dense sampling of the conduction band, we use Wannier interpolation based on the Wannier90 code \cite{Pizzi2019wannier90} with more than 2000 {\bf k}-points sampled along each high-symmetry path in the BZ.
To assess the effect of strain in (111)- and (100)-oriented ScN films, we constrain the in-plane lattice parameters and relaxed the out-of-plane lattice parameter.

The electronic structure calculations preceding phonon and electron mobility calculations are performed using the Quantum ESPRESSO software package \cite{giannozzi2009quantum, giannozzi2017advanced}.
We therefore perform calculations using the local density approximation (LDA) \cite{perdew1992accurate}.  In Sec.~\ref{sec:bands} we demonstrate, by comparing with the HSE calculations, that the relevant parts of the band structure are accurately reproduced with this approach.
The LDA calculations are performed with norm-conserving pseudopotentials \cite{troullier1991efficient} and the same valence electron configurations as specified above.
The lattice parameter is set to the experimental value of 4.50 \AA\ \cite{travaglini1986electronic}.
We used a plane-wave cutoff of 72 Ry and a 24 $\times$ 24 $\times$ 24 $k$-point grid to integrate over the BZ.

The phonon band structure is calculated using DFPT \cite{baroni2001phonons} with a threshold of 10$^{-14}$ Ry for self-consistency and using a coarse 6 $\times$ 6 $\times$ 6 $q$-point grid.
A non-analytical correction \cite{gonze1994interatomic} to the phonon bands has been included to capture the split in frequencies between longitudinal and transverse optical modes.
Based on the lattice vibrational spectra and electronic wavefunctions, we calculate el-ph coupling using the EPW code \cite{ponce2016epw}.
The el-ph matrix elements are first calculated on a coarse grid of 12 $\times$ 12 $\times$ 12 $k$- and 6 $\times$ 6 $\times$ 6 $q$-point grids, and then Wannier interpolation is performed at denser $k$- and $q$-point grids.
To calculate the scattering rates and mobility, we use 120 $\times$ 120 $\times$ 120 $k$- and $q$-point grids.
We have confirmed that our calculated mobility changes by less than 1\% when the grid size is increased to 140 $\times$ 140 $\times$ 140.
The Perturbo code \cite{zhou2020perturbo} was also employed for transport calculations, yielding consistent results for the mobility.

 \subsection{Electron-Phonon Matrix Elements} \label{e-phmatrix}

ScN has two atoms in its primitive unit cell, each of which carries an isotropic Born effective charge tensor.
The atomic motion associated with a longitudinal optical (LO) phonon in the long-wavelength limit ($\mathbf{q}\rightarrow0$) generates a long-range electric field that couples strongly to carriers \cite{ziman2001electrons}.
This is known as the Fr\"ohlich interaction \cite{frohlich1954electrons}.
As a consequence, LO-phonon scattering is the major scattering mechanism at finite temperatures, when LO phonon modes are thermally excited.

In the Fr\"ohlich model \cite{frohlich1954electrons}, the el-ph (LO) coupling matrix $g_\mathbf{q}$ is proportional to $1/\mathbf{q}$, diverging when $\mathbf{q}\rightarrow0$.
This singularity requires special numerical treatment in calculations for the el-ph interaction.
We have adopted the approach of Verdi \emph{et al.} \cite{verdi2015frohlich} to calculate the el-ph matrix elements by separating the short-range and long-range contributions to the coupling strength.
The el-ph matrix element $g_{mn,\nu}(\mathbf{k}, \mathbf{q})$ that quantifies the scattering process between Kohn-Sham states ($n$,$\mathbf{k}$) and ($m$,$\mathbf{k+q}$), interacting with a phonon with wavevector $\mathbf{q}$ and band index $\nu$, can be written as:
\begin{equation}\label{polar}
g_{mn,\nu} (\mathbf{k}, \mathbf{q}) = g^S_{mn,\nu} (\mathbf{k}, \mathbf{q}) + g^L_{mn,\nu} (\mathbf{k}, \mathbf{q}) ,
\end{equation}
$g^S_{mn,\nu}$ is the short-range contribution, which is obtained from Wannier interpolation on a dense grid \cite{verdi2015frohlich}.
$g^L_{mn,\nu} (\mathbf{k}, \mathbf{q})$ is the long-range contribution, which is computed analytically from
\begin{equation}\label{longrange}
\begin{split}
&g^L_{mn,\nu} (\mathbf{k}, \mathbf{q})  = \\
&i\frac{e^2}{\Omega \epsilon_0} \sum_j\sqrt{\frac{\hbar}{2M_j\omega_{\mathbf{q},\nu}}}
\sum_{\mathbf{G}\mathbf{q}} \frac{(\mathbf{q}+\mathbf{G})\cdot \mathbf{Z}^*_j \cdot e_{j,\nu}(\mathbf{q})}{(\mathbf{q}+\mathbf{G})\cdot \epsilon^{\infty}\cdot {(\mathbf{G}+\mathbf{q})}} \\
& \times  \left \langle  \psi_{m,\mathbf{k}+\mathbf{q}} \left | e^{i(\mathbf{q}+\mathbf{G})\cdot \mathbf{r}}  \right | \psi_{n,\mathbf{k}} \right \rangle .
\end{split}
\end{equation}
$\Omega$ is the volume of the cell, $\epsilon_0$ the permittivity of vacuum, $\mathbf{G}$ is the reciprocal lattice vector, and $\mathbf{Z}^*_j$ is the Born effective charge tensor of atom $j$ that has a mass $M_j$.
$\mathbf{e}_{j,\nu}(\mathbf{q})$ are the components of the eigenvector of the dynamical matrix for the phonon with band index $\nu$ and wavevector $\mathbf{q}$ for atom $j$.
$\epsilon^{\infty}$ is the macroscopic high-frequency dielectric tensor.
$\left \langle  \psi_{m,\mathbf{k}+\mathbf{q}} \left | e^{i(\mathbf{q}+\mathbf{G})\cdot \mathbf{r}}  \right | \psi_{n,\mathbf{k}} \right \rangle$ is the overlap matrix element between the initial and the final state and is evaluated in the limit of $\mathbf{q}+\mathbf{G}\rightarrow 0$.

Recent work highlighted the importance of dynamical quadrupoles in calculations
of carrier mobilities \cite{brunin2020electron,jhalani2020piezo}.
In this work we do not include quadrupole corrections because ScN crystallizes in the $Fm\bar 3 m$ space group, hence quadrupole tensors vanish by symmetry (see for example the similar case of SrO in \cite{ponce2021firstprinciples}).

In doped semiconductors, the el-ph interaction is screened by the free charge carriers.
We take this effect into account by screening the el-ph matrix elements using the Lindhard dielectric function, i.e., $g^\mathrm{screen}_{mn,\nu}(\mathbf{q}) = g_{mn,\nu}(\mathbf{q})/\epsilon(\mathbf{q})$.
The Lindhard function is evaluated at the phonon frequency $\omega_{{\bf q}\nu}$, and corresponds to the RPA screening of the homogeneous electron gas \cite{Lindhard,hedin1965new}.
The screened el-ph matrix elements  are then used in the calculation of the electron mobility \cite{caruso2016theory,verdi2017origin}.

\subsection{Electron Transport and Carrier Mobility}

Electron mobility is calculated as
\begin{equation}\label{eq:mob}
\mu_{\alpha \beta}  = -\frac{e}{\Omega n_e} \sum_{n \in \mathrm{CB}} \int \frac{d^3\mathbf{k}}{\Omega_{BZ}}  v_{n\mathbf{k},\alpha} \partial_{E_\beta}f_{n\mathbf k} ,
\end{equation} 
where $\Omega$ is the volume of the unit cell, $n_e$ is the electron concentration, $\Omega_{BZ}$ is the volume of the BZ, $f_{n,\mathbf{k}}$ is the Fermi-Dirac distribution function, and $v_{n\mathbf{k},\alpha}=\hbar^{-1} \partial \epsilon_{n\mathbf{k}}/{\partial k_{\alpha}}$  is the band velocity for the single-particle electron eigenvalue $\epsilon_{n\mathbf{k}}$.
One can define the band- and $k$-dependent el-ph relaxation time ($\tau_{n\mathbf{k}}$) that goes into the integral on the right side of Eq.~(\ref{eq:mob}).
The scattering rate ($\tau^{-1}_{n\mathbf{k}}$, defined as the inverse of the relaxation time) can be directly calculated from the imaginary part of the el-ph self-energy Im$\Sigma_{nk}$ using $\tau^{-1}_{n\mathbf{k}} =(2/\hbar) \mathrm{Im}\Sigma_{nk}$, which is expressed as
\begin{equation}\label{rates}
\begin{split}
\tau^{-1}_{n\mathbf{k}} & = \frac{2\pi}{\hbar} \sum_{\mathbf{q} \nu,m} |g_{mn,\nu}(\mathbf{k},\mathbf{q})|^2   \\
& \times \{(n_{\mathbf{q},\nu}+f_{\mathbf{k}+\mathbf{q},m})\delta(\epsilon_{\mathbf{k}+\mathbf{q},m}-\epsilon_{\mathbf{k},n}-\hbar \omega_{\mathbf{q},\nu}) \\
& + (1 + n_{\mathbf{q},\nu} -f_{\mathbf{k}+\mathbf{q},m} ) \delta(\epsilon_{\mathbf{k}+\mathbf{q},m} - \epsilon_{\mathbf{k},n} + \hbar \omega_{\mathbf{q},\nu} )\}  ,
 \end{split}
\end{equation}
where $\omega_{\mathbf{q},\nu}$ is the frequency and  $n_{\mathbf{q},\nu}$ the occupation (using Bose-Einstein statistics) of the phonon mode $\nu$ at wavevector $\mathbf{q}$.
The two Dirac delta functions indicate the energy conservation conditions for el-ph scattering events, with the former corresponding to a phonon absorption process and the latter to a phonon emission process.
The scattering rate is temperature dependent.
While the el-ph matrix is calculated at $T$=0, the temperature dependence of the scattering rate arises from the electron and phonon occupations at finite temperatures.
The electron mobility is obtained from an iterative solution of the Boltzmann transport equation \cite{ponce2018towards}.

\section{Results and Discussion} \label{sec:result}

\subsection{Details of Conduction-Band Structure} \label{sec:bands}

\begin{figure}
\includegraphics[scale=0.40]{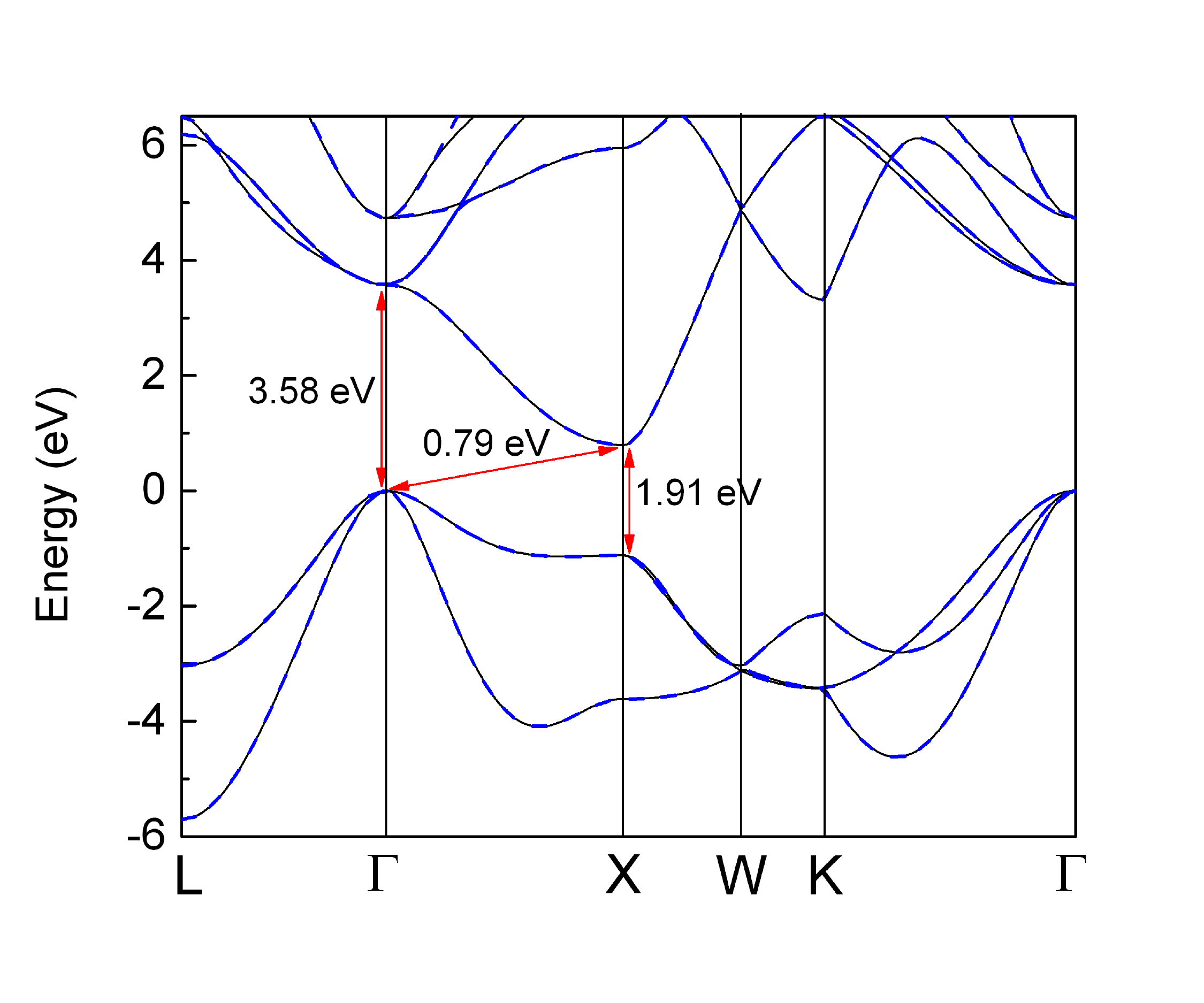}
\caption{The band structure of rocksalt ScN. Results from HSE calculations (black solid lines) and Wannier90 interpolation (blue dashed lines) are overlaid.  Indirect and direct band gaps are indicated.}
\label{fig:bulk}
\end{figure}

The band structure of rocksalt ScN as calculated with HSE is displayed in Fig.~\ref{fig:bulk}.
The VBM is located at the $\Gamma$ point and the CBM at the X point; there are six equivalent X points on the BZ boundary.
The valence band has O $2p$ character, while the conduction band is comprised of Sc $3d$ states.
The lowest three conduction bands belong to the unoccupied $t_{2g}$ orbitals, which arise from crystal-field splitting of Sc $3d$ orbitals in the octahedral ligand field.
We find an indirect band gap of 0.79 eV from $\Gamma$ to X, and optical direct gaps of 1.91 eV at X and 3.58 eV at $\Gamma$.
These values are in reasonable agreement with the range of computational \cite{saha2017compensation,qteish2006exact,stampfl2001electronic,lambrecht2000electronic} and experimental \cite{al2004surface, gall2001electronic,saha2013electronic,deng2015optical} results that have been previously reported.
Arguably the most detailed measurements were reported by Deng \emph{et al.} \cite{deng2015optical} who performed optical measurements that yielded direct gaps of 2.02 eV at X and 3.75 eV at $\Gamma$. Using scanning tunneling spectroscopy, Al-Brithen \emph{et al.} \cite{al2004surface} obtained an indirect band gap of 0.9$\pm$0.1 eV.
Our results are also in good agreement with the hybrid functional calculations of Deng \emph{et al.} \cite{deng2015optical} and with GW calculations \cite{qteish2006exact}.

In Fig.~\ref{fig:bulk} we overlay the Wannier-interpolated bands onto the band structure calculated with HSE, demonstrating that the Wannier fit is well converged.
Figure~\ref{fig:bulk} shows very different CB dispersion along the path connecting the X point (0, 0.5, 0.5) with $\Gamma$ (0, 0, 0) versus the path connecting X with W (0.25, 0.75, 0.5).
These high-symmetry pathways are perpendicular and are generally referred to as the longitudinal (X$\rightarrow \! \Gamma$) and transverse (X$\rightarrow$W) directions.
The secondary conduction-band valleys are considerably higher in energy than those at X; specifically, the K valley is 2.52 eV higher than the CBM at X, while the $\Gamma$ valley is 2.78 eV higher.
Therefore, electrons will only occupy the X valleys over a wide range of temperatures and carrier concentrations, and scattering to other valleys will not be a concern.

In this paper we focus on electron transport; however, in the course of investigating the HSE band structure, we have similarly analyzed the valence bands; we include our results for the valence bands in the Supplemental Material (SM) \cite{SM}.

In Fig.~\ref{fig:CB} we plot the band velocities and effective masses as a function of energy above the CBM.
We determine the band velocities by taking the first derivative with respect to {\bf{k}} on our Wannier-interpolated HSE band structure using the Wannier90 band interpolation tool \cite{Pizzi2019wannier90}.
The band velocity is defined as
\begin{equation}
  {v}_{\bf k} = \frac{1}{\hbar}\frac{\partial \epsilon_{\bf k}}{\partial {\bf k}} \, .
  \label{eq:velocity}
\end{equation}
We then perform finite-differences calculations on the first derivative to determine the second derivative with respect to {\bf{k}}, which is related to the effective mass tensor as:
\begin{equation}
\left[ \frac{1}{m^*_{\mathbf k}} \right]_{ij} = \frac{1}{\hbar^2} \frac{\partial^2\epsilon_{\mathbf k}}{\partial k_i \partial k_j} \, .
\end{equation}
In this way, we obtain accurate results for the evolution of effective mass as we move away from the CBM.

\begin{figure}
\includegraphics[scale=0.25]{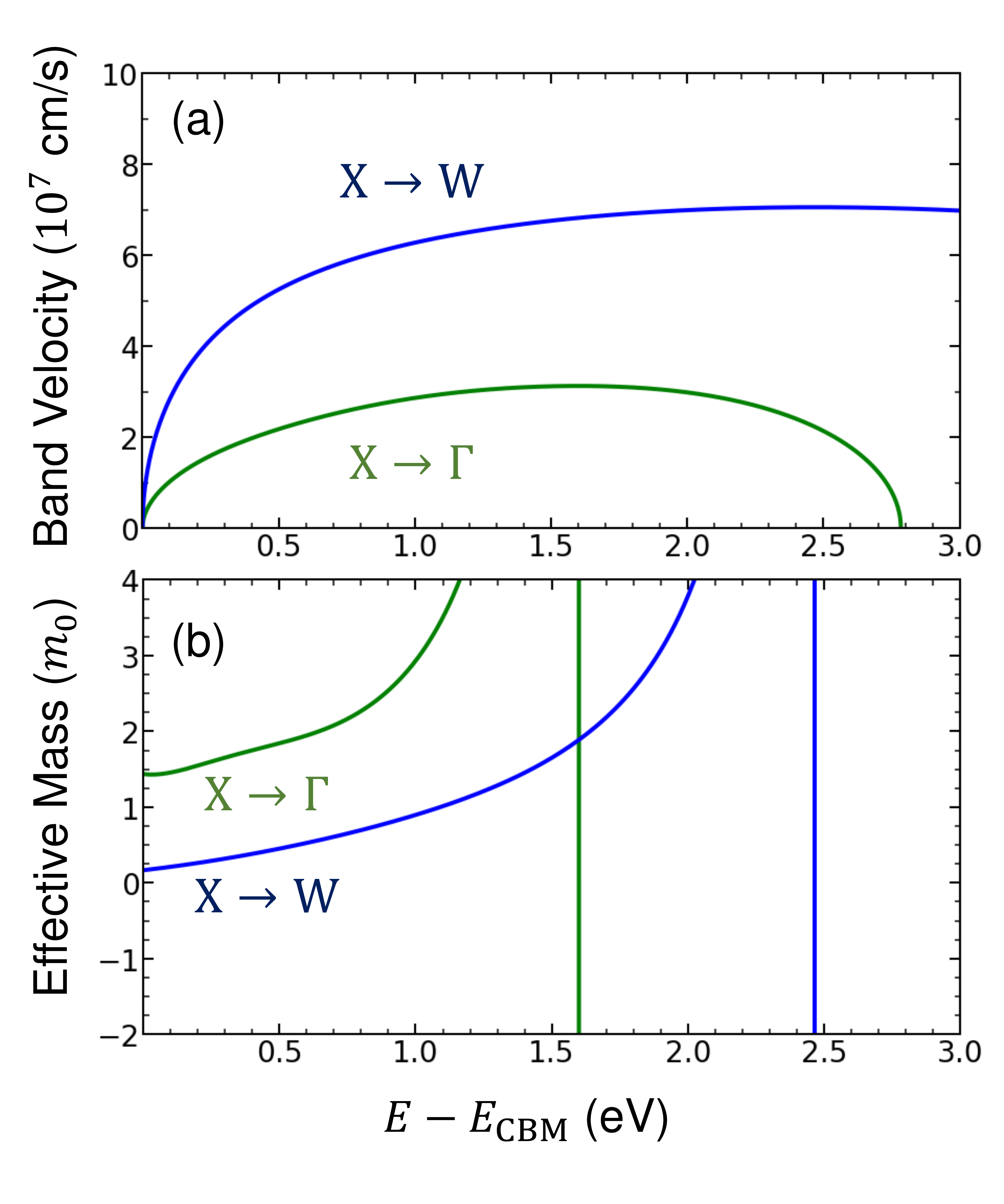}
\caption{(a) Band velocity and (b) effective mass as a function of energy above the CBM for the lowest-lying conduction band along X$\rightarrow$W and X$\rightarrow \! \Gamma$ directions. The vertical lines in (b) correspond to inflection points in the band structure, where the effective mass diverges.}
\label{fig:CB}
\end{figure}

The band velocity along the X$\rightarrow$W (transverse) direction is clearly greater, by more than a factor of two, than along the X$\rightarrow \! \Gamma$ (longitudinal) direction [Fig.~\ref{fig:CB}(a)].
This correlates with the significantly lower effective mass along X$\rightarrow$W [Fig.~\ref{fig:CB}(b)].
We can also locate the position of inflection points, i.e., energies for which the slope of the band velocity goes to zero, or, correspondingly, where the effective mass diverges.
The inflection point along X$\rightarrow \! \Gamma$ is approximately 1.6 eV above the CBM, while along X$\rightarrow$W it is almost 2.5 eV above the CBM.
If carriers are present at these inflection points, either by doping or by injection, they could be used in high-frequency oscillators \cite{kromer1958proposed,gribnikov2001negative,rowberg2020BSO}.

\begin{figure}
\includegraphics[scale=0.25]{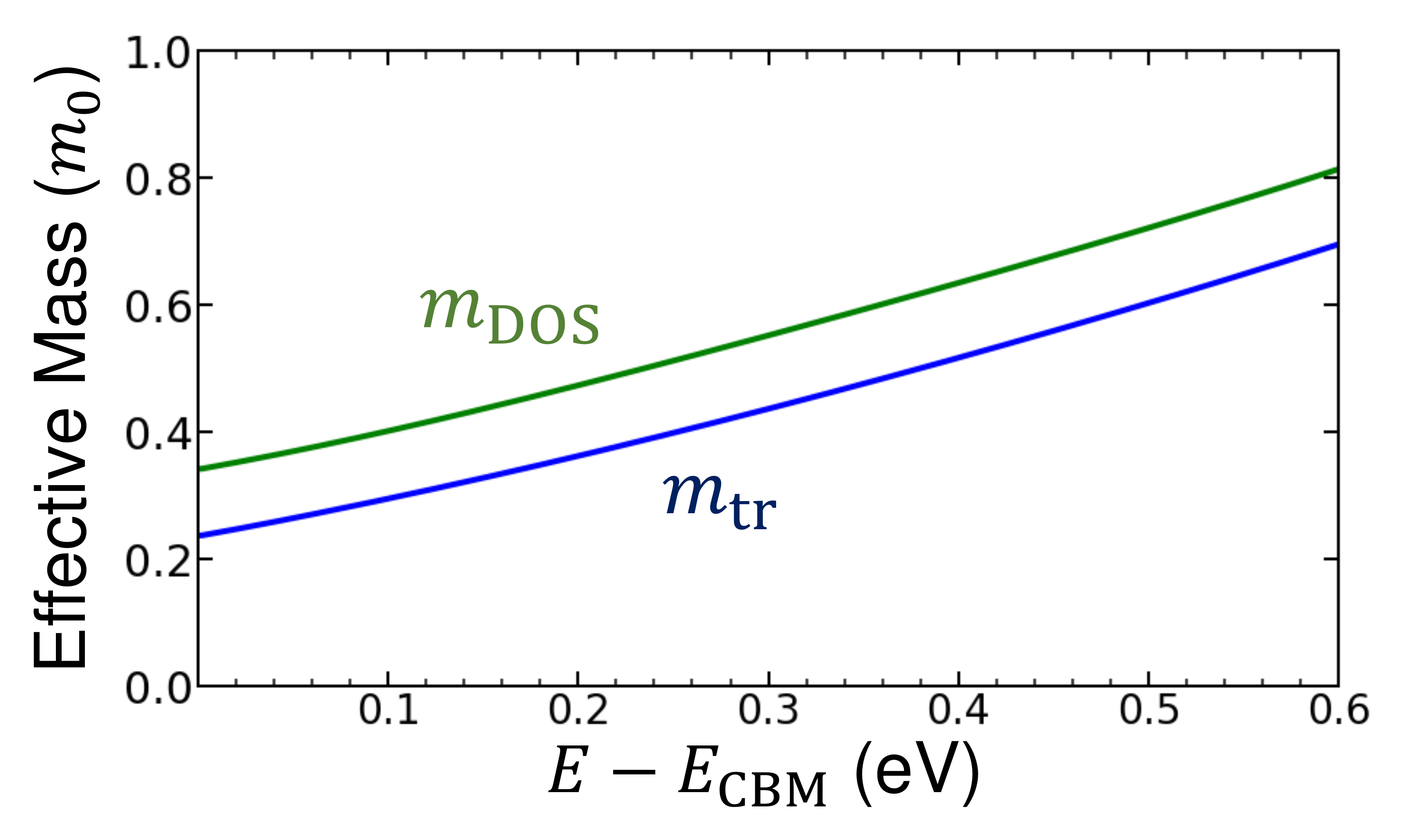}
\caption{Density-of-states and transport effective masses, shown as a function of electron energy above the CBM.
}
\label{fig:cbmass}
\end{figure}

Our calculated values for the longitudinal and transverse effective masses at the CBM are listed in Table~\ref{tab:mass}.
The values agree well with those derived from previous computational studies \cite{qteish2006exact,deng2015optical}.
We additionally calculate the transport effective mass ($m_\mathrm{tr}$)---sometimes also referred to as conductivity effective mass---and density-of-states effective mass ($m_\mathrm{DOS}$).
These are defined as
\begin{equation}
{m_\mathrm{tr}}=3\bigg(\frac{1}{m_{l}} + \frac{2}{m_{t}}\bigg)^{-1}
\label{eq:cond}
\end{equation}
and
\begin{equation}
{m_\mathrm{DOS}}={(m_{l}m^2_{t})}^{1/3} \, .
\label{eq:dos}
\end{equation}
The values as a function of energy above the CBM are shown in Fig.~\ref{fig:cbmass}.
As carriers are added into the conduction band, both masses increase, making clear that it is not warranted to assume that the masses retain the value they have at the CBM.

\begin{figure}
\includegraphics[scale=0.25]{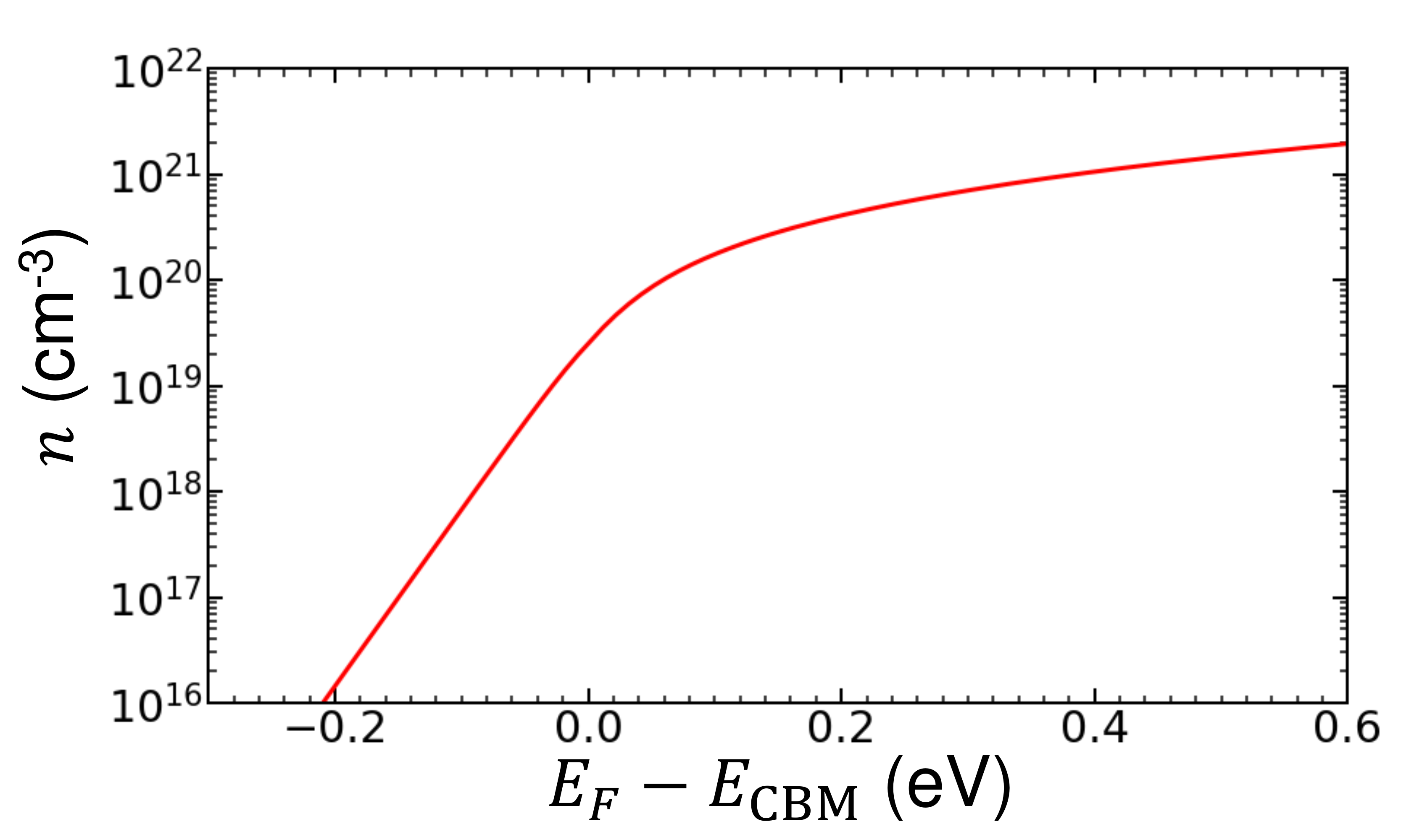}
\caption{Electron concentrations at $T$=300 K, shown as a function of the Fermi level position.
}
\label{fig:conc}
\end{figure}

The values for $m_{\mathrm{tr}}$ and $m_{\mathrm{DOS}}$ at the CBM are included in Table~\ref{tab:mass}.
Our value for $m_{\mathrm{tr}}$=0.24 $m_0$ is smaller than that reported by Deng $et$ $al.$ \cite{deng2015optical}, who extracted $m_{\mathrm{tr}}$=0.40$\pm$0.02 $m_0$ based on a fit of the plasma frequency that did not take the variation of $m_{\mathrm{tr}}$ with carrier concentration into account.
However, our calculated values are greater than those measured by Harbeke $et$ $al.$, who obtained effective masses of 0.1--0.2 $m_0$ using infrared reflectivity \cite{harbeke1972electron}.

For the density-of-states mass, we note that our definition does not include a factor $M_c^{2/3}$, where $M_c$ is the number of equivalent band minima ($M_c=3$ for the CBM in ScN) \cite{van2011principles}.
In this manner, we are following the definition used by Deng $et$ $al.$ in Ref.~\onlinecite{deng2015optical}, with whom our density-of-states effective mass agrees well.
They reported that $m_{\mathrm{DOS}}$=0.33$\pm$0.02 $m_0$ based on an analysis of the Burstein-Moss shift in optical absorption data \cite{deng2015optical}.

\begin{table}
\caption{Calculated electron effective masses $m_l$, $m_t$, $m_\mathrm{tr}$ and $m_\mathrm{DOS}$ (in units of $m_0$) using HSE and LDA band structures.  Reported experimental values (from Ref.~\onlinecite{deng2015optical}) are listed for comparison.}
\begin{ruledtabular}
\begin{tabular}{ccccc}
 	& $m_l$  & $m_t$ & $m_\mathrm{tr}$  & $m_\mathrm{DOS}$\\
\hline
HSE &  1.43    & 0.17     & 0.24    &  0.34  \\
LDA  & 1.67    & 0.19     & 0.27    &    0.39    \\
Expt$^a$  & 		& 	& 0.40$\pm$0.02          & 0.33$\pm$0.02 \\
\end{tabular}
\end{ruledtabular}
{$^a$Ref.~\onlinecite{deng2015optical}}
\label{tab:mass}
\end{table}

While the electronic structure computed with HSE is in good agreement with previous experimental and theoretical results,
we cannot use HSE for the calculation of mobility.  Indeed, we need the Quantum ESPRESSO code \cite{giannozzi2009quantum, giannozzi2017advanced} for the calculation of
electron-phonon matrix elements, but the DFPT implementation in this code is not compatible with HSE.
We therefore perform these calculations consistently with LDA, and demonstrate that the conduction-band structure that is relevant for electron mobility agrees well with the HSE band structure.
This agreement indicates that the band velocities and effective masses are reasonably accurate in LDA.
The underestimation of the band gap in LDA likely leads to a small amount of overscreening of the electron-phonon matrix elements, therefore the calculated mobility should be understood as a theoretical upper bound to the measured mobility~\cite{ponce2021firstprinciples}.
As a quantitative measure of the agreement between LDA and HSE results in the vicinity of the CBM, we compare electron effective masses calculated with LDA with those calculated with HSE, as listed in Table~\ref{tab:mass}.
The similarity in electron effective masses when comparing LDA with HSE assures us that using the LDA band structure is sufficient to yield reliable results for electron mobility.

\subsection{Electron Concentrations}
\label{sec:concen}

Using our Wannier-interpolated band structure and the corresponding density of states, we evaluate the relationship between carrier concentrations and Fermi level.
Electron concentrations $n$ are determined by integrating the density of states as a function of energy from the CBM \cite{van2011principles}:
\begin{equation}
	n=\int_{E_{CBM}}^\infty g_C(E)f(E)dE \, .
\label{eq:elconc}
\end{equation}
Here, $g_C(E)$ is the density of states in the conduction band as a function of energy, and $f(E)$ is the Fermi-Dirac occupation function, given by:
\begin{equation}
	f(E)=\frac{1}{1+\exp(\frac{E-E_F}{k_BT})} \, ,
\label{eq:Fermi}
\end{equation}
where $k_B$ is Boltzmann's constant and $E_F$ is the Fermi energy.
In Fig.~\ref{fig:conc}, we plot $n$ at room temperature as a function of Fermi level, obtained through numerical integration.
A similar analysis for holes in the valence band is included in the Supplemental Material \cite{SM}.

Deng \emph{et al.} \cite{deng2015optical} reported optical band gaps as a function of increasing electron concentration starting at $1.0\times10^{20}$ cm$^{-3}$.
The observed an increase in optical band gap, caused by a Burstein-Moss shift, by 0.37 eV when increasing the carrier concentration from $1.0\times10^{20}$ cm$^{-3}$ to $1.0\times10^{21}$ cm$^{-3}$.
If we assume a flat valence band around the X point, as Deng \emph{et al.} did when fitting their data, we can compare this increase with the expected shift in Fermi level shown in Fig.~\ref{fig:conc}.
We find that increasing $n$ from $1.0\times10^{20}$ cm$^{-3}$ to $1.0\times10^{21}$ cm$^{-3}$ corresponds to a shift of 0.33 eV, in very good agreement with the results of Deng \emph{et al.} \cite{deng2015optical}.
We also obtain excellent agreement with the results of Gall \emph{et al.} \cite{gall2001electronic}, who estimated a Burstein-Moss shift of 0.25 eV for a carrier concentration of $5.0\times10^{20}$ cm$^{-3}$.
Our results indicate that the same concentration will increase the position of the Fermi level by 0.23 eV.

\subsection{Phonon Band Structure of ScN} \label{sec:phononbands}

\begin{figure}
\includegraphics[width=0.5\textwidth]{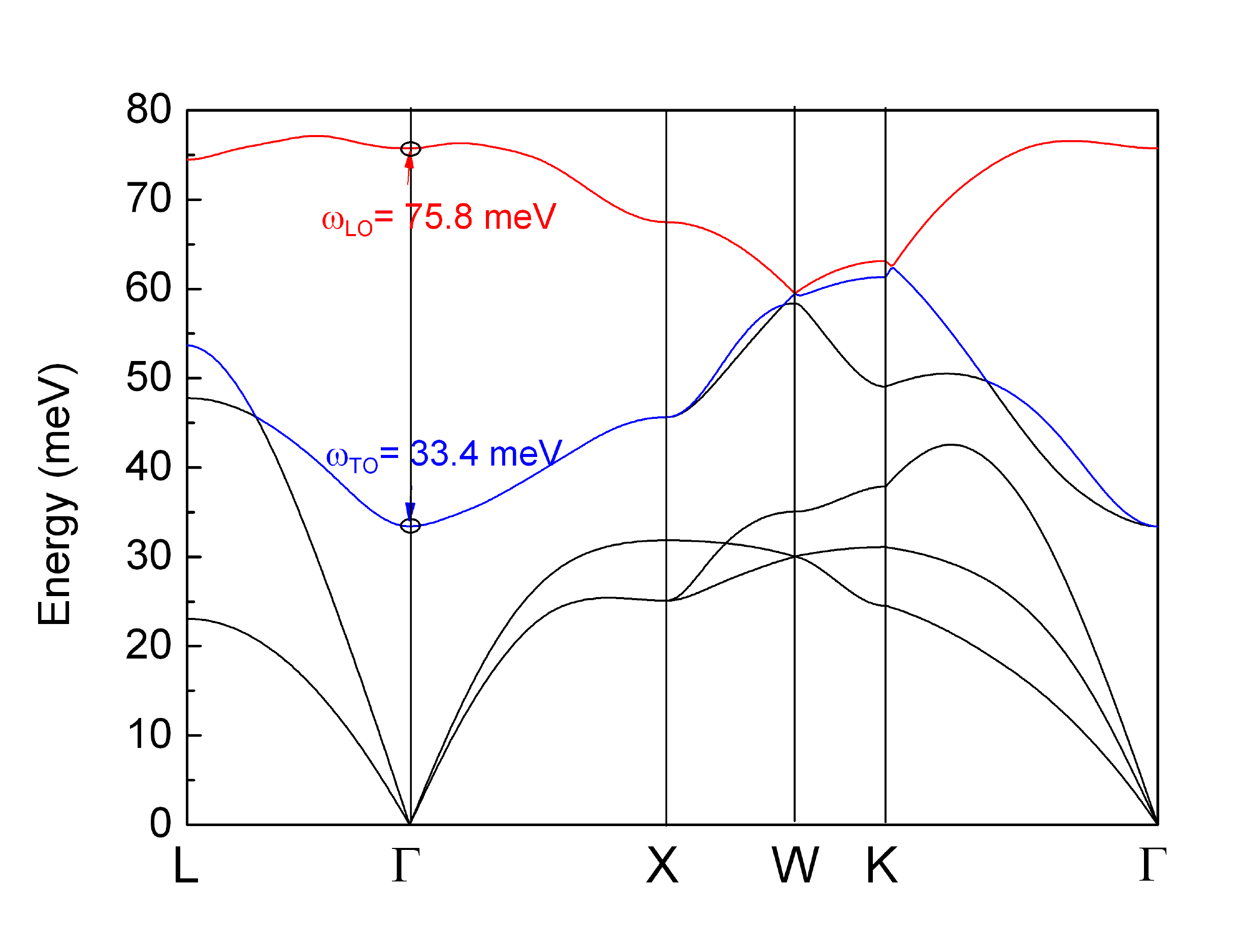}
\caption{\label{fig:phononbands}
Calculated phonon band structure along high-symmetry lines in the BZ. Frequencies for the LO and TO phonon modes at the $\Gamma$ point are indicated. }
\end{figure}

Our calculated phonon band structure for ScN is shown in Fig.~\ref{fig:phononbands}.
With two atoms in the primitive cell, ScN possesses six phonon modes in total: one longitudinal acoustic (LA) mode,  two transverse acoustic (TA) modes, one longitudinal optical (LO) mode, and two transverse optical (TO) modes.
Our calculated phonon frequencies at $\Gamma$ and X are listed in Table~\ref{tab:omega}.
The calculated phonon frequencies generally agree with those from a previous LDA calculation~\cite{paudel2009calculated}, except that $\omega_\mathrm{TO}(\Gamma)$ in the present study is 11.9 meV lower.
This discrepancy may be attributed to the different pseudopotentials and codes being used.
Experimentally measured phonon frequencies \cite{uchiyama2018phonon,travaglini1986electronic,gall2001vibrational} are also listed in Table~\ref{tab:omega}.
While our calculated phonon frequencies at the X point are in reasonable agreement with experiment \cite{uchiyama2018phonon,travaglini1986electronic, gall2001vibrational}, the optical phonon frequencies at the $\Gamma$ point are underestimated.
However, we note that the values obtained from different experiments [inelastic X-ray scattering (IXS)~\cite{uchiyama2018phonon} versus Raman spectroscopy~\cite{travaglini1986electronic, gall2001vibrational}] also differ substantially.
Our calculated LO-TO splitting [$\omega_\mathrm{LO}(\Gamma)-\omega_\mathrm{TO}(\Gamma)=42.3$ meV] is consistent with the IXS measurement (42.6 meV).
This value arises from the nonanalytical correction at $\mathbf{q}\rightarrow0$ and requires both the high-frequency dielectric constant ($\epsilon^{\infty}$) and the diagonal term of the isotropic Born effective charge tensor, which we will call $Z^*$.
Our calculations yield $\epsilon^{\infty}$=9.8 and $Z^*$(Sc)=$-Z^*$(N)=3.8 $e$.
These values are comparable to those calculated with GGA+$U$ in Ref.~\onlinecite{saha2010electronic}: $\epsilon^{\infty}$=12.9 and $Z^*$(Sc)=4.4 $e$.
The origin of the differences between results calculated at different levels of theory is well-known and has previously been discussed~\cite{dalcorso1994dft}. 
In our own HSE calculations we found $\epsilon^{\infty}$=8.4 and $Z^*$(Sc)=4.1 $e$, in general agreement with the LDA-calculated values.

\begin{table}
\caption{Calculated LO, TO, LA, TA phonon frequencies (in meV) for ScN at the $\Gamma$ and X points.
Results from a previous calculation~\cite{paudel2009calculated} are listed for comparison.
Experimental results from inelastic X-ray scattering (IXS)~\cite{uchiyama2018phonon} and Raman spectroscopy~\cite{travaglini1986electronic, gall2001vibrational} are also listed.}
\begin{ruledtabular}
\begin{tabular}{c|cc|cccc}
    & \multicolumn{2}{c|}{$\Gamma$} &\multicolumn{4}{c}{X}  \\
\hline
    & $\omega_\mathrm{LO}$ &  $\omega_\mathrm{TO}$ & $\omega_\mathrm{LO}$& $\omega_\mathrm{TO}$& $\omega_\mathrm{LA}$ & $\omega_\mathrm{TA}$  \\
\hline
This work  &75.8 & 33.4  & 69.3 & 50.7 & 42.8 & 32.8   \\
Prev. calc.$^a$  & 78.4 & 45.3 & 70.5 & 51.5 & 43.9 &33.1 \\
IXS$^b$  & 85.1 & 42.5  & 67.5 & 45.7 & 31.8 & 25.1 \\
Raman$^c$  & 84.3 &52.1 & -- & 52.0 & 44.0 & 37.0 \\
\end{tabular}
\end{ruledtabular}
{$^a$Ref.~\onlinecite{paudel2009calculated}; $^b$Ref.~\onlinecite{uchiyama2018phonon}; $^c$Ref.~\onlinecite{travaglini1986electronic, gall2001vibrational}}
\label{tab:omega}
\end{table}

\subsection{Electron-Phonon Matrix Elements} \label{sec:matrix}

\begin{figure}
\includegraphics[width=0.50\textwidth]{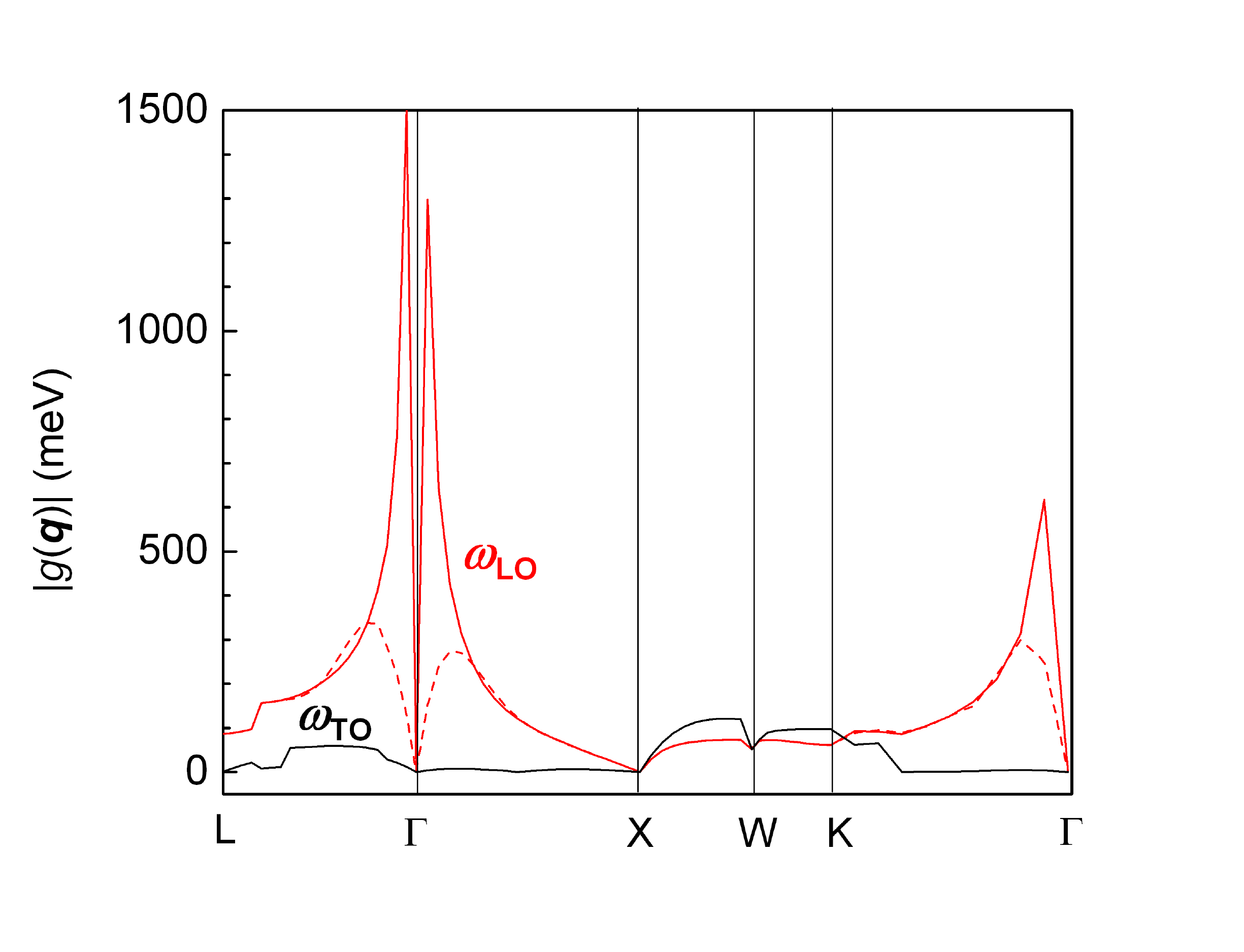}
\caption{\label{fig:matrix}
Magnitude of el-ph matrix elements $|g_{mn,\nu} (\mathbf{k}, \mathbf{q})|$ for electrons in the lowest CB ($m=n=5$, $\mathbf{k}$=X) and for the highest phonon branch (LO mode, $\nu=6$, red solid line) and the second-highest phonon branch (TO mode, $\nu=5$, black solid line).
The red dashed line shows calculated values for the short-range contribution $|g^S_{mn,\nu} (\mathbf{k}, \mathbf{q})|$.
}
\end{figure}

In Fig.~\ref{fig:matrix}, we present our results for el-ph matrix elements $|g_{mn,\nu} (\mathbf{k}, \mathbf{q})|$ between two electron states in the lowest conduction band ($m=n=5$, $k=$X) and an optical phonon branch $\nu$.
We consider the intraband electron interaction with either the polar LO mode or the nonpolar TO mode, and the resulting $|g_{mn,\nu} (\mathbf{k}, \mathbf{q})|)$ is plotted as a function of phonon wavevector $\mathbf{q}$ along the high-symmetry paths of the BZ.
The el-ph matrix elements for the polar LO mode are clearly much larger than those for the TO mode.
For the electron-LO phonon interaction, a strong $1/\mathbf{q}$ dependence is observed for small $\mathbf{q}$, due to the long-range contribution $|g^L_{mn,\nu} (\mathbf{k}, \mathbf{q})|$ [see Eq.~(\ref{longrange})].
The importance of this long-range contribution is highlighted by a comparison with the calculated values for the short-range contribution $|g^S_{mn,\nu} (\mathbf{k}, \mathbf{q})|$, shown as the dashed red line in Fig.~\ref{fig:matrix}.
Regarding acoustic modes, we found that the el-ph matrix elements for the LA mode are comparable in magnitude to those for the TO mode, while those for the TA mode are negligibly small.

\subsection{Scattering Rate} \label{sec:rates}
\begin{figure}
\includegraphics[width=0.49\textwidth]{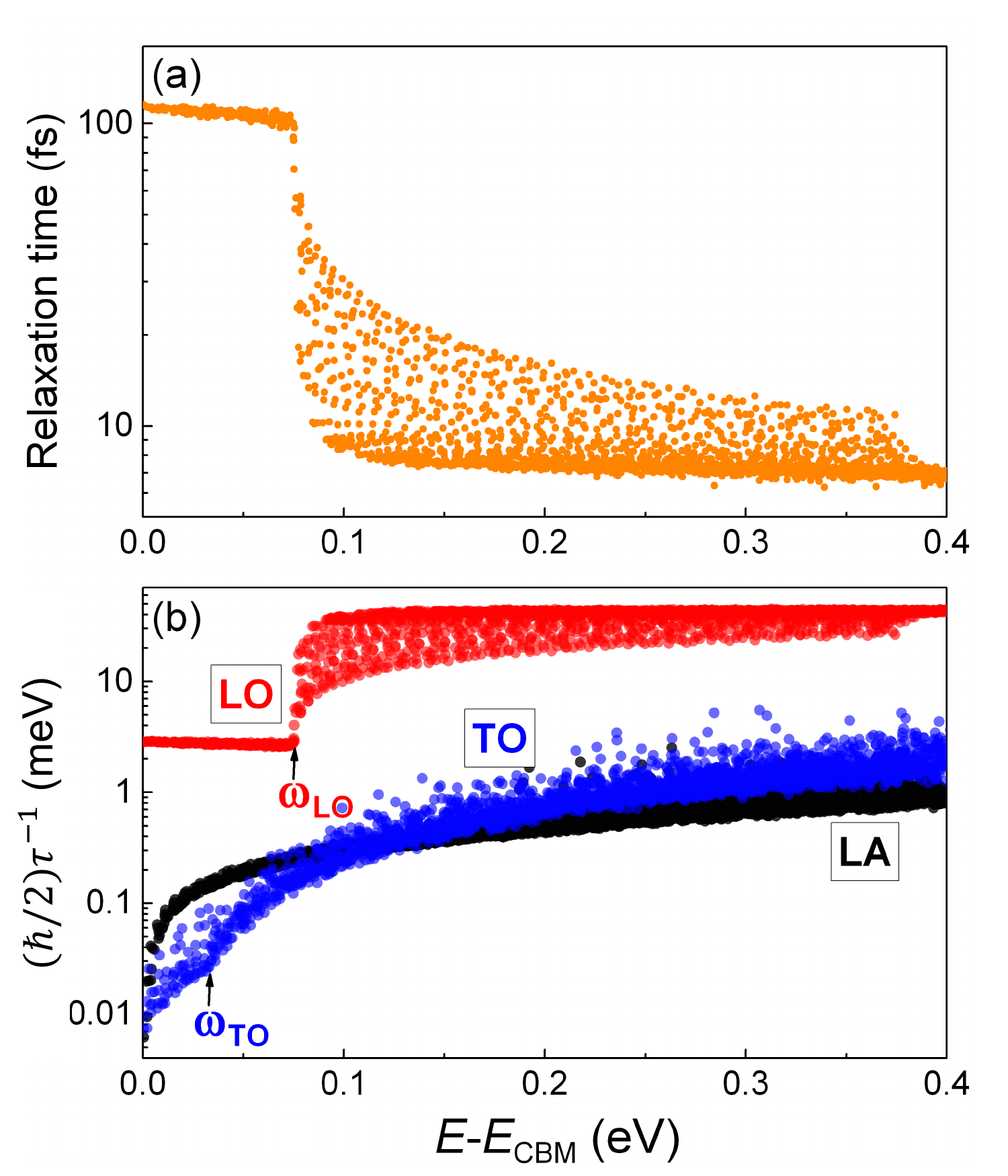}
\caption{\label{fig:scatter}
(a) Relaxation time of electrons ($\tau$, fs).
(b) Mode-resolved imaginary part of the electron self-energy (lm$\Sigma$, meV), which is proportional to the scattering rate $\tau^{-1}$ with a proportionality factor $(\hbar/2)$. Contributions from the dominant LO, TO, and LA modes are shown, and the characteristic frequencies $\omega_\mathrm{LO}$ and $\omega_\mathrm{TO}$ are indicated.}
\end{figure}

Using the calculated el-ph matrix elements, we evaluate the room-temperature relaxation times [Fig.~\ref{fig:scatter}(a)] for electrons in the conduction band [see Eq.~(\ref{rates})].
The results are shown for energies up to 0.4 eV above the CBM.
As the satellite valleys (K and $\Gamma$ valleys) are much higher in energy than the X valley (see Fig.~\ref{fig:bulk}), only small-$\mathbf{q}$ intravalley electron scattering within the X valley is possible.
The relaxation time decreases slightly over the energy range 0--76 meV, ranging from 104 to 114 fs.
Above 76 meV, the relaxation time exhibits a sharp decrease,
reaching a plateau around 7--8 fs at higher electron energies.

To gain more insight into el-ph scattering, we plot the mode-resolved el-ph scattering rates in Fig.~\ref{fig:scatter}(b).
As expected, and consistent with the mode-resolved el-ph matrix elements shown in Fig.~\ref{fig:matrix}, scattering with LO phonons dominates.
The total scattering rate mainly follows the trend of LO scattering, which exhibits a threshold frequency of $\omega_\mathrm{LO} (\Gamma)=76$ meV.
At energies less than $\omega_\mathrm{LO}$ from the CBM, the phase space for LO phonon emission vanishes, leaving LO phonon absorption as the sole source of scattering events.
At energies above  $\omega_\mathrm{LO}$, both LO phonon emission and absorption are possible, sharply increasing the scattering rate.

Compared to LO phonon scattering, TO phonon scattering is much weaker.
TO phonon scattering also has a threshold frequency (at $\omega_\mathrm{TO}$) below which phonon emission is suppressed and the scattering rate is lower.
Threshold frequencies in scattering rates are commonly observed in inelastic electron--optical phonon scattering processes \cite{lundstrom2009fundamentals}.

Scattering with acoustic phonons, on the other hand, is nearly elastic, and its rate increases with carrier energy without any threshold frequency~\cite{lundstrom2009fundamentals}.
Overall, the LA phonon scattering rates are comparable to those of TO phonons at high carrier energies, but they are noticeably higher at low carrier energies.

\subsection{Carrier Mobility}\label{sec:mob}

\begin{figure}
\includegraphics[width=0.49\textwidth]{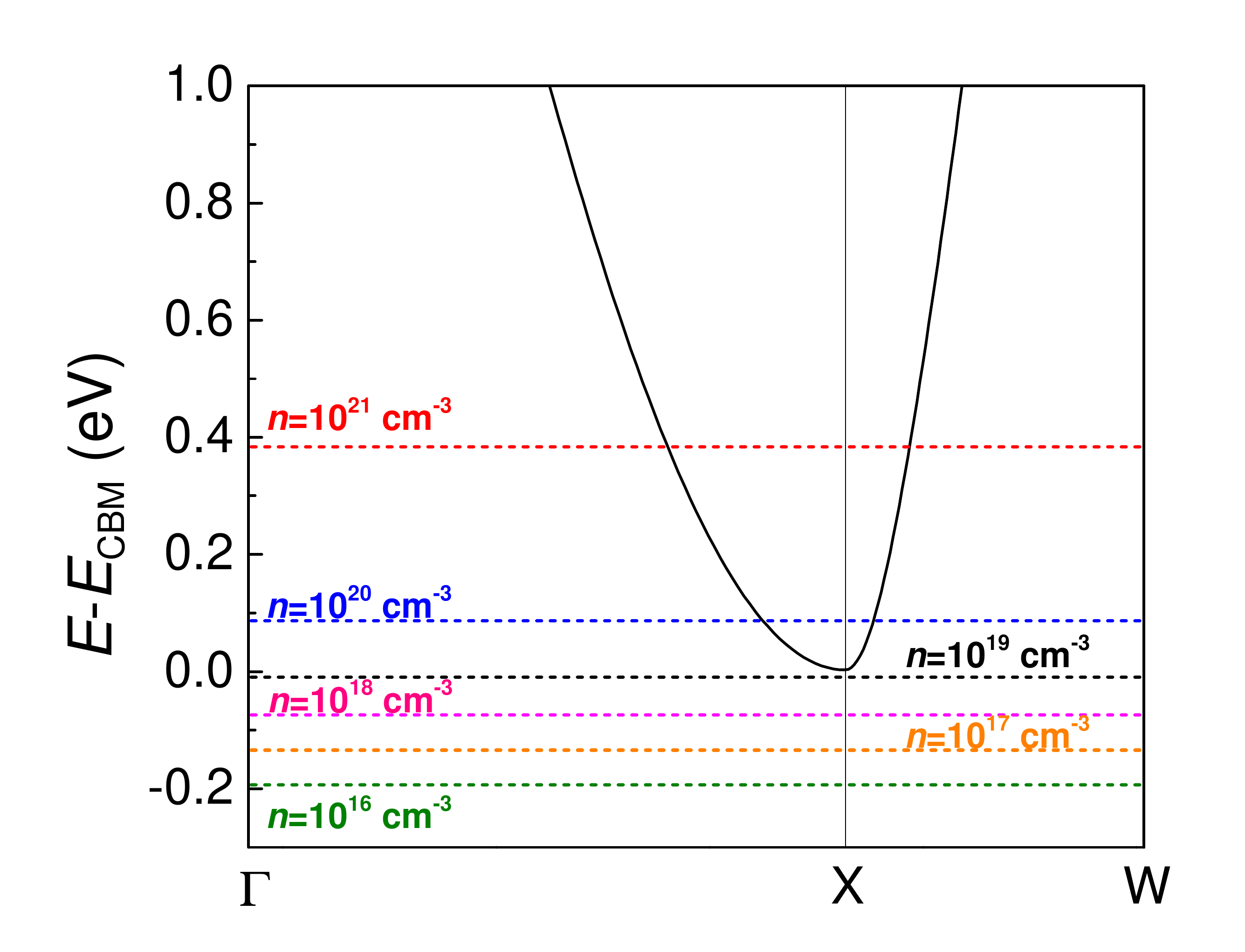}
\caption{\label{fig:mob-a}
Position of the Fermi level for different electron concentrations at room temperature. The zero of energy is set at the CBM.}
\end{figure}

With our calculated scattering rates, we now compute the room-temperature electron mobility of ScN using Eq.~(\ref{eq:mob}).
Since the energy scale for phonons (up to about 76 meV, Fig.~\ref{fig:phononbands}) is much smaller than for electrons, and because the factor $-\frac{\partial f_{n\mathbf{k}}}{\partial \epsilon_{n\mathbf{k}}}$ is nonzero only over a narrow range near the Fermi level, only states with energies close to the Fermi level are involved in scattering.
To compute the room-temperature mobility we focus on an energy window of 0.8 eV centered at the Fermi level.
We considered electron concentrations ranging from $10^{16}$ cm$^{-3}$ to $10^{21}$ cm$^{-3}$.
The Fermi level that corresponds to a particular carrier concentration has been labeled in Fig.~\ref{fig:mob-a}.
At $n=10^{16}$ cm$^{-3}$, the Fermi level is 0.19 eV below the CBM, while at $n=10^{19}$ cm$^{-3}$ it is positioned 0.01 eV below the CBM.
At our highest carrier concentration considered, $n=10^{21}$ cm$^{-3}$, the Fermi level is 0.38 eV above the CBM.

\begin{figure}
\includegraphics[width=0.5\textwidth]{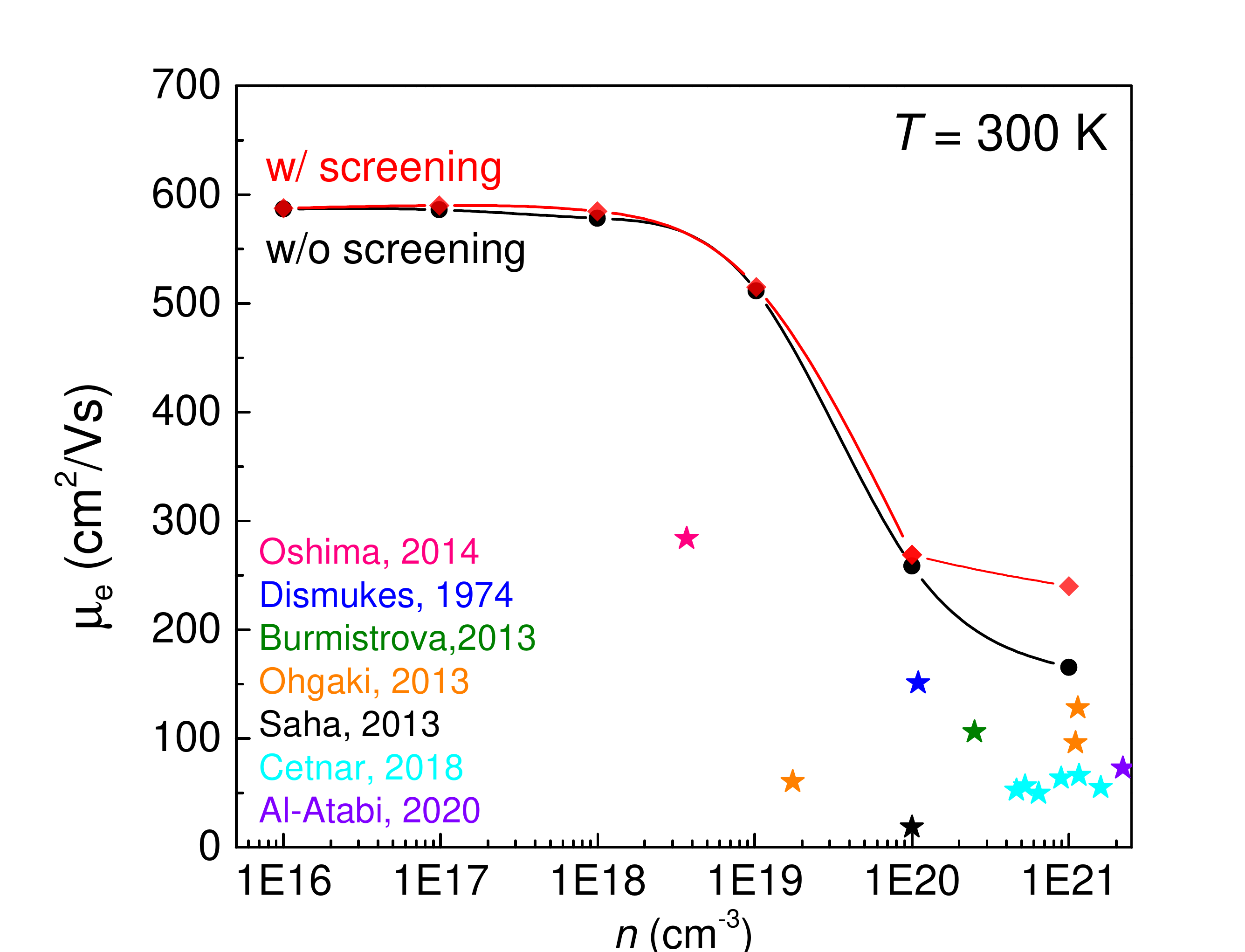}
\caption{\label{fig:mob-b}
Calculated room-temperature mobilities (in cm$^2$/Vs) for ScN as a function of electron concentration. The calculated mobilities with and without the screening effect are depicted using red and black dots, respectively.  Experimental values (denoted by stars) are from
Oshima {\it et al.}~\cite{oshima2014hydride}, Dismukes {\it et al.}~\cite{dismukes1972epitaxial}, Burmistrova {\it et al.}~\cite{burmistrova2013thermoelectric}, Ohgaki {\it et al.}~\cite{ohgaki2013electrical}, Saha {\it et al.}~\cite{saha2013electronic}, Cetnar {\it et al.}~\cite{cetnar2018electronic}  and Al-Atabi {\it et al.}~\cite{al2020properties}.}
\end{figure}

In Fig.~\ref{fig:mob-b}, we plot the room-temperature electron mobility as a function of the electron concentration.
When $n<10^{18}$ cm$^{-3}$, the electron mobility is relatively insensitive to $n$.
The electron mobility then drops quickly for higher
When $n$ increases above $10^{18}$ cm$^{-3}$, the mobility starts dropping quickly, from 578 cm$^2$/Vs at $n=10^{18}$ cm$^{-3}$ to 165 cm$^2$/Vs at $n=10^{21}$ cm$^{-3}$.

\begin{figure}
\includegraphics[width=0.48\textwidth]{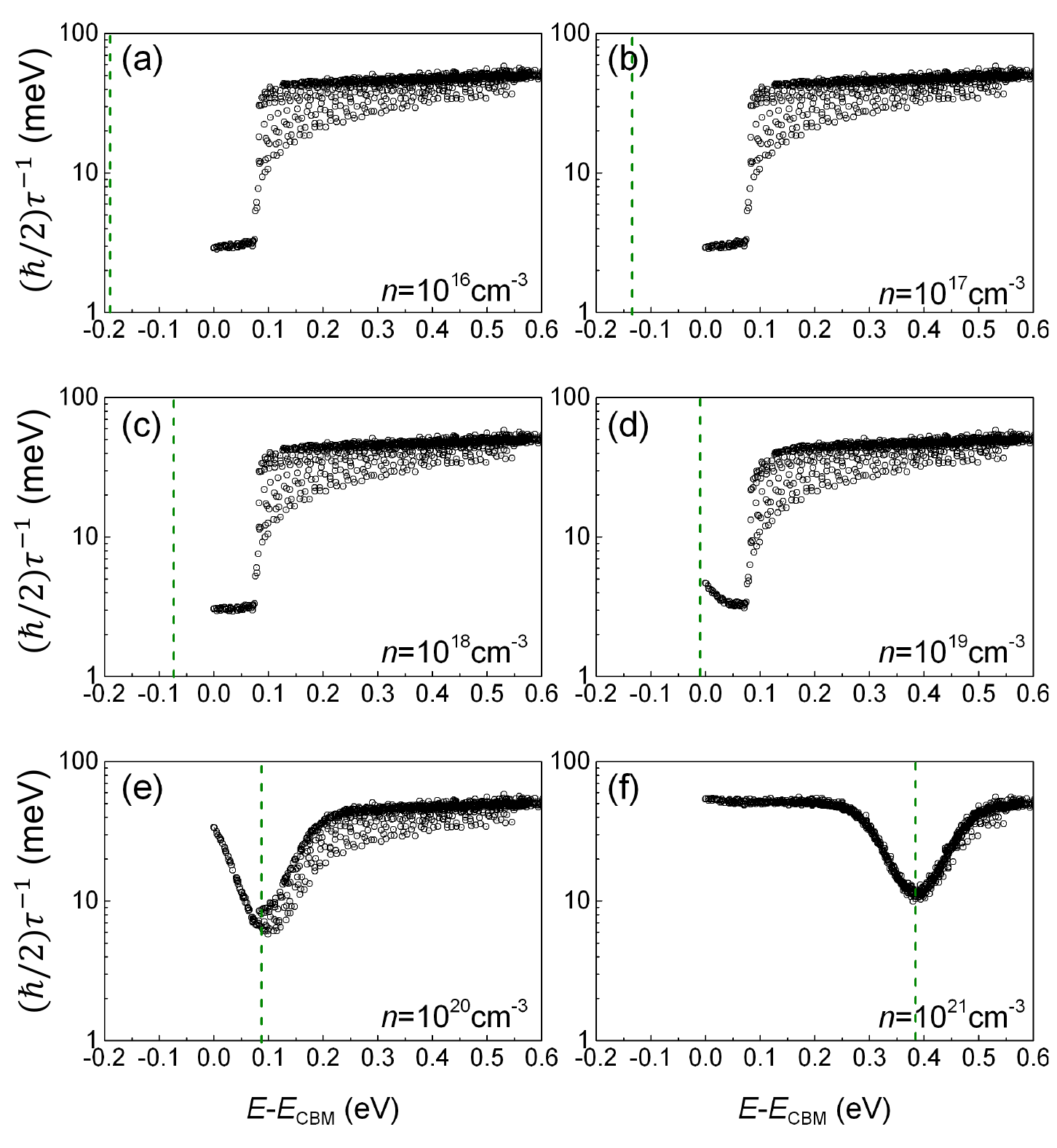}
\caption{\label{fig:conc_rates}
Room-temperature scattering rates as a function of carrier energy at different carrier concentrations. The zero of energy is set at the CBM. The Fermi-level position is labeled with green dashed lines.}
\end{figure}

To gain more insight into these results, we plot the total scattering rates versus electron energy for six different electron concentrations in Fig.~\ref{fig:conc_rates}.
For $n=10^{16}$, $10^{17}$, and $10^{18}$ cm$^{-3}$ the Fermi level is well below the CBM and the shape and magnitude of the scattering rate are similar.
At $n=10^{19}$ cm$^{-3}$ the Fermi level approaches the CBM and the scattering rate near the CBM starts to rise, because more electronic states participate in electron scattering [see the first term in Eq.~(\ref{rates})].
When the Fermi level is well above the CBM, for $n=10^{20}$ cm$^{-3}$, the scattering rates for low-energy carriers near the CBM are greatly increased.
The rates saturate at $n=10^{21}$ cm$^{-3}$.

For large carrier concentrations, there is always a dip in the scattering rate at the Fermi level.
We have verified from mode-resolved contributions to the scattering rates that this dip is caused by electron--LO phonon scattering.
The dip in the scattering rate around the Fermi energy originates from the fact that electron scattering by optical phonons is forbidden below the TO/LO energy (see also Ref.~\onlinecite{krishnaswamy2017first}.
The reduced scattering rate at the Fermi level is also consistent with the fact that electrons near the Fermi surface of a \textit{metal} are scattered less compared to electrons far from the Fermi surface.
We note that only the scattering rates near the Fermi level (within $\sim$0.2 eV of $E_F$, at room temperature) affect the electron mobility.
The scattering rate near the Fermi level gradually increases when the carrier concentration is increased above $n=10^{19}$ cm$^{-3}$ [Figs.~\ref{fig:scatter}(b) and \ref{fig:conc_rates}].
This trend stems from the increasing electron density of states and the associated phase space for scattering as the Fermi level moves up inside the conduction band, thus explaining the reduction in electron mobility.

\subsection{Screening Effect on Electron Transport}
\label{sec:screen}

Inclusion of free-carrier screening reduces the el-ph matrix elements and thereby increases the relaxation time and enhances the mobilities, as shown by the results in Fig.~\ref{fig:mob-b}.
To identify the carrier concentration at which screening is expected to play an notable role~\cite{caruso2016theory,verdi2017origin}, we estimate the plasma frequency $\omega_\mathrm{P}$ of free electrons and compare it to the frequency of LO phonons, which dominate the electron-phonon scattering.
When $\omega_\mathrm{P} < \omega_\mathrm{LO}$, free-carrier screening is ineffective \cite{verdi2017origin}; we expect it to start affecting the electron-phonon interaction when $\omega_\mathrm{P} > \omega_\mathrm{LO}$.
We calculate $\omega_\mathrm{P}$ as a function of carrier concentration $n$ based on the standard expression~\cite{lundstrom2009fundamentals}
\begin{equation}
\label{eq:plasma}
\omega_\mathrm{P}=\sqrt{\frac{n e^2}{m^*\epsilon_0 \epsilon^{\infty}}} \, .
\end{equation}
$\omega_\mathrm{P}$ is 48 meV at $n=10^{19}$ cm$^{-3}$, and increases to 150 meV at $n=10^{20}$ cm$^{-3}$; comparing these values with
the LO phonon frequency ($\omega_\mathrm{LO}=76$ meV), we expect free-carrier screening to become important around $n=10^{20}$ cm$^{-3}$.
At $n$=$10^{21}$ cm$^{-3}$ we find $\omega_\mathrm{P}$=330 meV and el-ph coupling is notably suppressed, resulting in a signficant mobility enhancement (Fig.~\ref{fig:mob-b})---from 165 cm$^2$/Vs to 240 cm$^2$/Vs, an increase by 45\%.

Figure~\ref{fig:mob-b} also contains values from various experiments \cite{dismukes1972epitaxial,oshima2014hydride,burmistrova2013thermoelectric,ohgaki2013electrical, saha2013electronic}.
Our calculated mobility values are systematically higher.
This is to be expected, because we only include electron-phonon scattering; our values thus constitute an upper bound on the mobility.
Other scattering mechanisms, such as ionized impurity scattering, are present in actual samples.
We also note that our calculations are for the drift mobility, while some of the experimental data~\cite{dismukes1972epitaxial,ohgaki2013electrical,cetnar2018electronic} are for the Hall mobility.
Recent benchmarks on standard semiconductors indicate that the calculated electron Hall mobility is typically lower than the drift mobility \cite{ponce2021firstprinciples}.

\subsection{Strain Effects on Mobility} \label{sec:strain}

Strain engineering is an established strategy for tailoring carrier mobilities in semiconductors. 
Different substrates will lead to different strains, either as residual strains in a partially relaxed layer, or as a result of pseudomorphic growth of thin films.
We investigated application of in-plane strain (denoted by $\varepsilon_{\parallel}$) in a (111) plane or in a (100) plane. The former may occur for epitaxial growth on AlN or GaN, while the latter would occur as residual strain when ScN is grown on Si or MgO.

\begin{figure}
\includegraphics[width=0.49\textwidth]{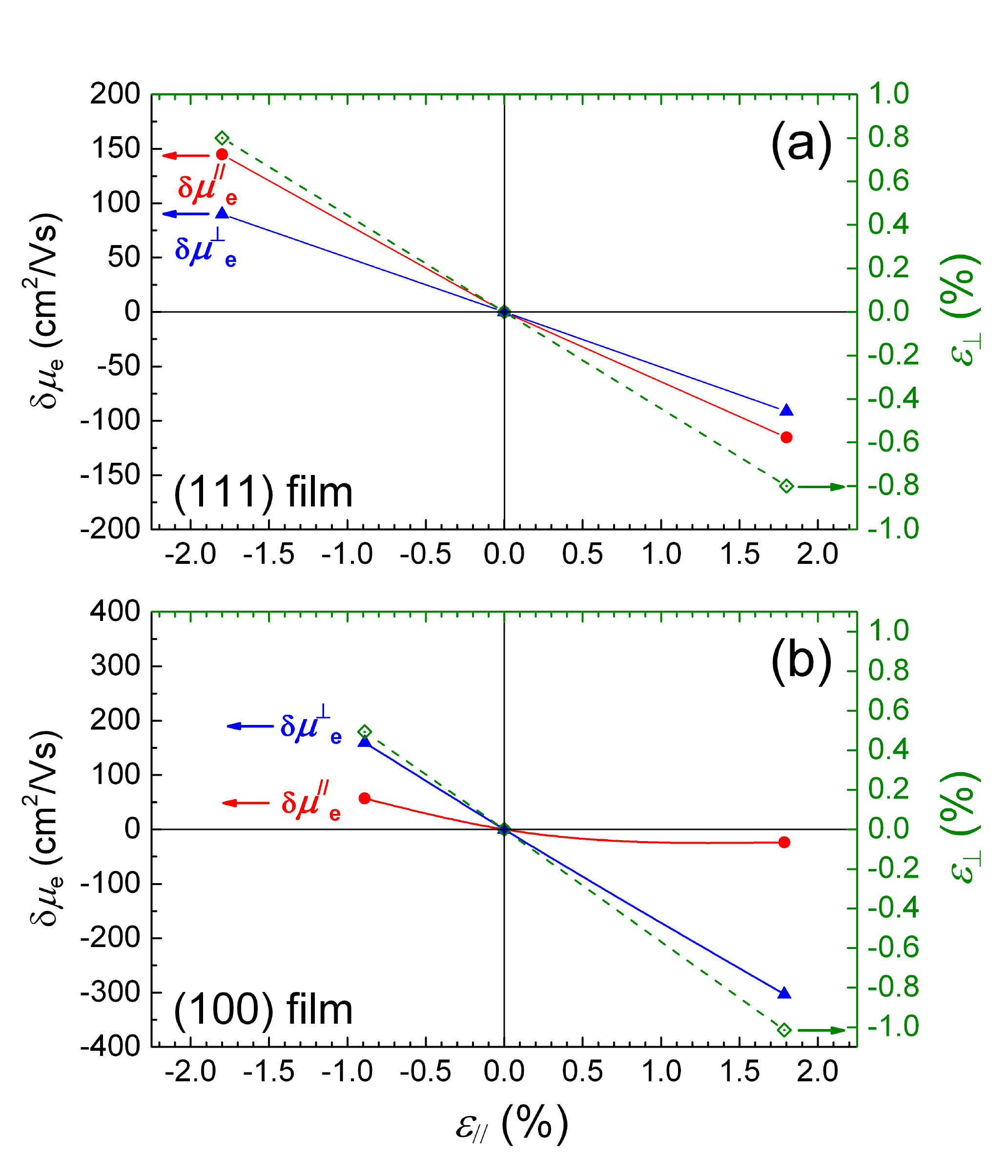}
\caption{\label{fig:strain}
Calculated change in room-temperature mobility ($\delta \mu_e$, left axis) and relaxed out-of-plane strain ($\varepsilon_\perp$, right axis) under different in-plane strains ($\varepsilon_\parallel$) in (a) (111)-oriented ScN films and (b) (100)-oriented ScN films. $\delta \mu^{\parallel}_e$ denotes the variation of in-plane mobility while $\delta \mu^{\perp}_e$ denotes the variation of out-of-plane mobility.}
\end{figure}

The in-plane strain $\varepsilon_{\parallel}$ is accompanied by an out-of-plane lattice relaxation, described by $\varepsilon_{\perp}$, along either the [111] direction or the [100] direction.
For the [111]-oriented films, our HSE calculations show that 1.0\% compressive (tensile) epitaxial strain yields 0.44\% tensile (compressive) out-of-plane strain $\varepsilon_{\perp}$ [Fig.~\ref{fig:strain}(a)].
For the [100]-oriented films, 1.0\% compressive (tensile) epitaxial strain yields 0.56\% tensile (compressive) out-of-plane strain [Fig.~\ref{fig:strain}(b)].
Based on the strain relationships in Fig.~\ref{fig:strain}, we obtain a Poisson's ratio of 0.180 for [111]--oriented films and 0.219 for [100]-oriented films.
Both values compare well with the experimental measurements: 0.188 for [111]~\cite{moram2006young} and 0.201 for [100]~\cite{gall1999microstructural}.

We evaluated the change in the room-temperature electron mobility at $n=10^{16}$ cm$^{-3}$ as a function of in-plane strain.
Figure~\ref{fig:strain}(a) shows our results for (111) films.
Epitaxial strain in the (111) plane does not lift the degeneracy of the CBM at the X point;
it only alters the band curvature of the conduction band and hence the electron effective mass and band velocity.
We see that compressive strain increases the mobility.
This can be attributed to an increase in the dispersion of the conduction band.
The mobility along the out-of-plane direction ($\mu^{\perp}_e$) follows the same trend as the in-plane mobility, but with a lower rate of change.
Tensile strain leads to a change of the same magnitude in the opposite direction; the behavior is close to linear over this range of strains.
A linear fitting shows that a 1.0\% compressive epitaxial strain increases the in-plane mobility by 72 cm$^2$/Vs and the out-of-plane mobility by 50 cm$^2$/Vs.

The results for (100) films are displayed in Fig.~\ref{fig:strain}(a).
Epitaxial strain in the (100) plane splits the threefold-degenerate conduction band at the X point (not including spin degeneracy) into a singly-degenerate and a doubly-degenerate band.
Under compressive in-plane strain, the singly-degenerate band is higher in energy. The splitting between the singly-degenerate and doubly-degenerate band extrema is 36 meV per 1\% epitaxial strain.
The out-of-plane mobility ($\mu^{\perp}_e$) displays a strong dependence on strain: a 1\% compressive in-plane strain raises the mobility by 172 cm$^2$/Vs.
The in-plane mobility ($\mu^{\parallel}_e$) is not very sensitive to strain, and exhibits a notable nonlinearity.
This can be attributed to the lifting of the degeneracy at the X points, which impacts electron occupation and scattering.

\section{Conclusion} \label{sec:conc}

In summary, we reported an in-depth first-principles study of the conduction-band structure, electron-phonon interaction and room-temperature electron transport in ScN.
We determined band velocities and effective masses as a function of carrier concentration.
Electron mobilities were calculated as a function of carrier concentrations ranging from $10^{16}$ to $10^{21}$ cm$^{-3}$.
Effects of free-carrier screening were included; we found that screening significantly enhances electron mobility when the carrier concentration exceeds 10$^{20}$ cm$^{-3}$.
We calculated a room-temperature electron mobility of 587 cm$^2$/Vs at $n$=10$^{16}$ cm$^{-3}$, decreasing to 240 cm$^2$/Vs at $n$=10$^{21}$ cm$^{-3}$.
We also showed that strain can strongly impact the electron mobility.
In (111)-oriented ScN films a 1.0\% compressive strain increases the in-plane mobility by 72 cm$^2$/Vs and the out-of-plane mobility by 50 cm$^2$/Vs.
The increase in out-of-plane mobility is even greater in (100)-oriented ScN films: a 1.0\% compressive strain increases the out-of-plane mobility by 172 cm$^2$; the impact of (100) strain on the in-plane mobility is quite weak.
These results demonstrate how pseudomorphic strain can potentially be used to enhance the electron mobility.
In light of the critical importance of high mobility for ScN's applications, including in thermoelectric and (opto)electronic devices, our results will help support its further technological development.

\begin{acknowledgments}
We are grateful to John Cetnar, David Look, Mark Turiansky, Nicholas Adamski, Karthik Krishnaswamy, Wennie Wang, and Youngho Kang for valuable discussions, to
Jinjian Zhou for assistance with the Perturbo code, and to Samuel Ponc\'e for useful discussions about the EPW code.
This work was supported by the Air Force Office of Scientific Research under award number FA9550-18-1-0237.
A.J.E.R. was also supported by the National Science Foundation (NSF) Graduate Research Fellowship Program under Grant No. 1650114. Any opinions, findings, and conclusions or recommendations expressed in this material are those of the author(s) and do not necessarily reflect the views of the NSF.
J.L. and F.G.  acknowledge support from the Computational Materials Sciences Program funded by the U.S.  Department of Energy, Office of Science, Basic Energy Sciences, under Award DE-SC0020129.
We acknowledge computational resources provided by the Extreme Science and Engineering Discovery Environment (XSEDE), which is supported by NSF Grant No. ACI-1548562,
and by the DOD High Performance Computing Modernization Program at  the  AFRL  DSRC  and  ERDC  DSRC  under  Project  No. AFOSR46403464.
Use was also made of computational facilities purchased with funds from NSF (CNS-1725797) and administered by the Center for Scientific Computing (CSC). The CSC is supported by the California NanoSystems Institute and the Materials Research Science and Engineering Center (NSF DMR-1720256) at UC Santa Barbara.

\end{acknowledgments}

\bibliography{ScN_transport_051321}

\end{document}